\documentclass[aps,twocolumn,showpacs,pra,groupedaddress,superscriptaddress,nofootinbib,longbibliography]{revtex4-1}
\usepackage{graphicx,amsmath,amssymb,amsbsy,subfigure,hyperref,bbm,times,txfonts}
\usepackage{color}

\providecommand{\openone}{\leavevmode\hbox{\small1\kern-3.8pt\normalsize1}}

\begin{document}

\title{Validity of Landauer principle and quantum memory effects via collision models}

\author{Zhong-Xiao Man}
\email{zxman@qfnu.edu.cn}
\affiliation{Shandong Provincial Key Laboratory of Laser Polarization and Information Technology, Department of Physics, Qufu Normal University, Qufu 273165, China}

\author{Yun-Jie Xia}
\email{yjxia@qfnu.edu.cn}
\affiliation{Shandong Provincial Key Laboratory of Laser Polarization and Information Technology, Department of Physics, Qufu Normal University, Qufu 273165, China}

\author{Rosario Lo Franco}
\email{rosario.lofranco@unipa.it}
\affiliation{Dipartimento di Energia, Ingegneria dell'Informazione e Modelli Matematici, Universit\`{a} di Palermo, Viale delle Scienze, Edificio 9, 90128 Palermo, Italy}
\affiliation{Dipartimento di Fisica e Chimica, Universit\`a di Palermo, via Archirafi 36, 90123 Palermo, Italy}

\begin{abstract}
We study the validity of Landauer principle in the non-Markovian regime by means of collision models where the intracollisions inside the reservoir cause memory effects generating system-environment correlations. We adopt the system-environment correlations created during the dynamical process to assess the effect of non-Markovianity on the Landauer principle. Exploiting an exact equality for the entropy change of the system, we find the condition for the violation of the Landauer principle, which occurs when the established system-environment correlations become larger than the entropy production of the system. We then generalize the study to the non-equilibrium situation where the system is surrounded by many reservoirs at different temperatures. Our results, verified through collision models with Heisenberg-type interactions, suggest that the complexity of the environment does not play a significant role in the qualitative mechanisms underlying the violation of the Landauer principle under non-Markovian processes.
\end{abstract}


\maketitle

\section{Introduction}
The increasing abilities in the fabrication and characterization
of nanoscale systems allow us to explore the fundamental
laws of classical thermodynamics at the quantum level, which
gives birth to the subject of quantum thermodynamics \cite{QT}.
In the usual formulation of quantum thermodynamics,
one typically makes Markovian approximations for an open system interacting with ideal heat baths.
In this case, the environment is assumed to be able to relax to the equilibrium state at a time
scale much faster than that of the system, which induces a monotonic one-way flow of information from the system to the environment. However, in many practical situations
the Markov approximation is no longer valid \cite{RevModPhys.88.021002,RevModPhys.89.015001,lofrancoreview,RLFchapter2017}, so that non-Markovian (memory) effects on the dynamical features of the quantum system have to be taken into account.
Although considerable attention has been paid to the formulation of quantitative
measures of non-Markovianity \cite{measure1,measure2,measure3,measure4,measure5,measure6}, to its experimental demonstration \cite{NMexp1,NMexp2,NMexp3,NMexp4,NMexp5,XuNatCom}, and to the exploration of its origin \cite{origin1,origin2}, its role in affecting the thermodynamics properties of open quantum systems has so far remained little explored \cite{MultiLandauer,Implications,NMPower,cascaded,EbackNM,NequiLand}.
Effect of non-Markovianity in logically irreversible processes has been investigated based on the Landauer principle \cite{MultiLandauer,Implications}, which gives a direct link between information theory and thermodynamics at both classical and quantum level \cite{Land1,Land2,Land3,DavidNJP,jiangLandauer}.
It is known that the validity of Landauer principle breaks down in the case of non-Markovian environments \cite{MultiLandauer,Implications}.
By means of Landauer principle, it has been shown that the memory effects can control the amount of work extraction by erasure in the presence of realistic environments \cite{NMPower}. A non-exponential time behavior of heat flow in non-Markovian reservoir has been studied \cite{cascaded}, where a bipartite system interacts dissipatively with a thermal reservoir in a cascaded fashion.

An efficient microscopic framework in simulating the non-Markovian dynamics is the \textit{collision} model \cite{Rau1963,colli1,colli2,colli3,colli4,colli5,colli6,colli7,colli8,colli9,colli10,colli11,colli12,colli13,colli14,colli15,
colli16,colli17,colli18,colli19,ManXiaLo,cicca2017}, which has also been recently used in the context of quantum thermodynamics \cite{cascaded,MultiLandauer,Implications,colli21,colli22}. The collision model assumes that the environment is a chain of $N$ ancillas and the system of interest $S$ interacts, or collides, with an ancilla at each time step.
It has been shown that, when there are no initial correlations among the ancillas and no correlations are created between them during the process, a Lindblad master equation can be derived \cite{colli1,colli2}. Collision models allow to recover the dynamics of non-Markovian (indivisible) channels by setting suitable system-environment conditions, for instance by introducing either correlations in the initial state of the ancillas or ancilla-ancilla collisions in the interval between two system-ancilla collisions. In other words, the non-Markovian dynamics can be achieved in the collision model when the system-environment interaction is mediated by the ancillary degrees of freedom. The collision models are also relevant to study some fundamental physical systems, such as the emission of an atom into a leaky cavity with a Lorentzian, or multi-Lorentzian, spectral density or a qubit subject to random telegraph noise \cite{colli13}. The highly stable and configurable non-Markovian collision-based dynamics can be experimentally implemented with all-optical setups \cite{collexp1,collexp2}.

Albeit the Landauer principle has been shown to be violated in non-Markovian dynamics \cite{MultiLandauer,Implications}, the underlying mechanisms under this phenomenon remains to be understood. Here we address this aspect, elucidating the behavior of this thermodynamic principle in the context of quantum memory effects. In particular, we exploit collision models to study, in the non-Markovian regime, the validity of Landauer principle, finding the condition for its violation.
In the employed models, the information of system $S$ is firstly transferred to an ancilla of the reservoir via a system-ancilla collision, a part of which then goes to the nearest neighbor ancilla via the ancilla-ancilla intracollision so that the lost information of $S$ can be recovered at the next system-ancilla collision. The reservoir intracollisions lead to system-reservoir correlations, which allow us to assess the influence of non-Markovianity on the Landauer principle.
Following the framework of Ref.~\cite{enpro}, we first provide an expression for the entropy change of the system which, apart from the original reversible and irreversible terms, exhibits an additional contribution due to the established system-reservoir correlations in terms of mutual information.
We connect this expression to the Landauer principle and obtain the condition for its violation, which occurs when the established system-reservoir correlations are larger than the entropy production of the system. We moreover extend the study to the non-equilibrium case where the system is coupled with several different reservoirs. Also in this case, we show that the amount of correlations arising among the system and all the reservoirs during the dynamics determine whether he Landauer principle is satisfied or not. All these results are confirmed by means of collision models with Heisenberg-like coherent interactions.

The outline of this paper is the following. In Sec. \ref{sec2}, focusing on a quantum system coupled to a single reservoir, we present the condition for the violation of Landauer principle in the non-Markovian process and its verification via a physical (collision) model. In Sec. \ref{sec3}, we study the non-equilibrium situation in which the system is surrounded by many reservoirs at different temperatures. We obtain the condition for the violation of Landauer principle also in this case and demonstrate the result via a simple model involving two reservoirs. Finally, in Sec. \ref{sec4} we draw our conclusions.

\section{Landauer principle in a single non-Markovian reservoir}\label{sec2}

\subsection{Condition for the violation of Landauer principle}
We consider the information erasing process of a quantum system $S$ in contact with a reservoir consisting of a chain of $N$ identical ancillary qubits $R_{1},R_{2},\ldots,R_{N}$. The system $S$ and a generic reservoir qubit $R$ are described, respectively, by the Hamiltonians ($\hbar=1$)
\begin{equation}\label{HSR}
\hat{H}_{S}=\omega_{S} \hat{\sigma}_{z}^{S}/2,\quad
\hat{H}_{R}=\omega_{R} \hat{\sigma}_{z}^{R}/2,
\end{equation}
where $\hat{\sigma}_{z}^{\mu}=\left|1\right\rangle_{\mu}\left\langle1\right|-\left|0\right\rangle_{\mu}\left\langle0\right|$ is the Pauli operator and $\{\left|0\right\rangle_{\mu},\left|1\right\rangle_{\mu}\}$ are the logical states of the qubit $\mu=S,$ $R$ with transition frequency $\omega_{\mu}$ (hereafter, for simplicity, we take $\omega_{R}=\omega_S = \omega$).
Here, we focus on a non-Markovian reservoir through the introduction of interactions between two nearest-neighbor reservoir qubits $R_{n}$ and $R_{n+1}$ ($n=1,2, \ldots, N-1$).

\begin{figure}[t!]
\begin{center}
{\includegraphics[width=0.47\textwidth]{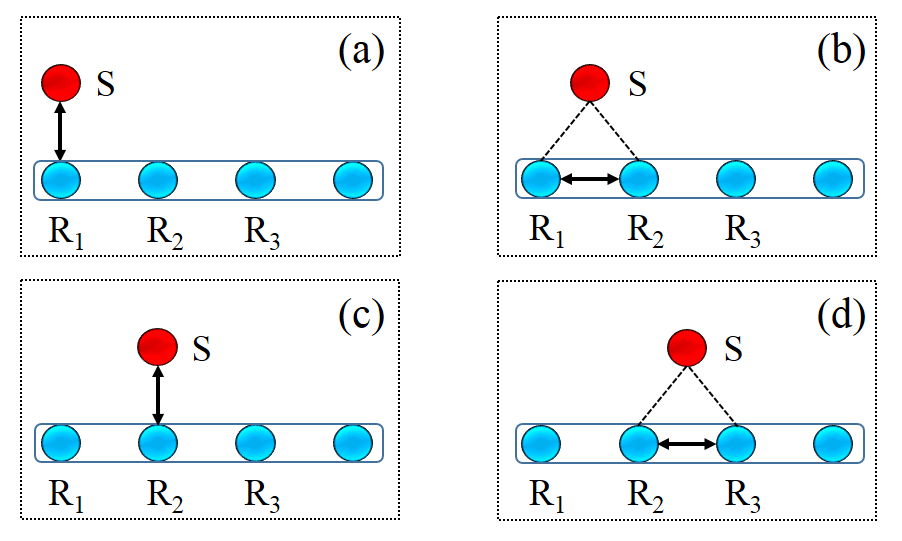}}
\end{center}
\caption{(Color online) (a) The system $S$ collides with the reservoir qubit $R_{1}$ which become correlated to each other, as denoted by the dashed line in the figure. (b) The intracollision between $R_{1}$ and $R_{2}$ takes place leading to the correlations among $S,$ $R_{1}$ and $R_{2}$. (c) The ancilla qubit $R_{1}$ is traced out, while $S$ and $R_{2}$ are still correlated. Then, $S$ collides with $R_{2}$, which is followed by intracollision of $R_{2}$-$R_{3}$ in panel (d). The process is then iterated accordingly.}
\label{M1}
\end{figure}

Our model is illustrated in Fig.~\ref{M1} for the first two steps of collisions. After the interaction between $S$ and $R_{1}$, the intra-reservoir collision between $R_{1}$ and $R_{2}$ occurs and then the system shifts forward one site to interact with $R_{2}$. Therefore, the correlation between $S$ and $R_{2}$ has been established prior to their collision. Generally, the bipartite state $\rho_{SR_{n}}$ of $S$-$R_{n}$ is correlated before their collision for $n\geq 2$ and such correlation is characterized by the mutual information $I(\rho_{SR_{n}})=S(\rho_{S,n})+S(\rho_{R_{n}})-S(\rho_{SR_{n}})$, where $\rho_{S,n}=\mathrm{Tr}_{R_{n}}\rho_{SR_{n}}$ ($\rho_{R_{n}}=\mathrm{Tr}_{S}\rho_{SR_{n}}$) is the marginal state of $S$ ($R_{n}$) and $S(\rho)=-\mathrm{Tr}(\rho\mathrm{ln}\rho)$ is the von Neumann entropy of $\rho$. The interaction between $S$ and $R_{n}$ is ruled by a unitary $\hat{U}_{SR_{n}}$, by which the state $\rho_{SR_{n}}$ is transformed to $\rho_{SR_{n}}^{\prime}=\hat{U}_{SR_{n}}\rho_{SR_{n}}\hat{U}_{SR_{n}}^{\dag}$. In the following, we label the marginal states of $S$ and $R_{n}$ after the collision as $\rho_{S_{n}}^{\prime}=\mathrm{Tr}_{R_{n}}\rho_{SR_{n}}^{\prime}$ and $\rho_{R_{n}}^{\prime}=\mathrm{Tr}_{S}\rho_{SR_{n}}^{\prime}$.
It is known that the total von Neumann entropy of $S$-$R_{n}$ under the unitary evolution is invariant, whether the state $\rho_{SR_{n}}$ is product or correlated: $S(\rho_{SR_{n}})=S(\rho_{SR_{n}}^{\prime})$.
In contrast, the entropy of the system is in general a function of time,
whose change $\Delta S_{n}$ by the interaction reads (see Appendix \ref{appenA} for details)
\begin{eqnarray} \label{DS}
  \Delta S_{n} &=& S(\rho_{S_{n}}^{\prime})-S(\rho_{S_{n}}) \nonumber\\
   &=&  D\left(\rho_{SR_{n}}^{\prime}\parallel\rho_{S_{n}}^{\prime}\rho_{R_{n}}\right)
  +\mathrm{Tr}_{R_{n}}(\rho_{R_{n}}^{\prime}-\rho_{R_{n}})\ln\rho_{R_{n}}\nonumber\\
  &&-I(\rho_{SR_{n}}),
\end{eqnarray}
where $D\left(\rho_{SR_{n}}^{\prime}\parallel\rho_{S_{n}}^{\prime}\rho_{R_{n}}\right)$ is the entropy production of the system representing the irreversible contribution to the entropy change of the system \cite{enpro}. Here, $D\left(\rho\parallel\rho^{\prime}\right)\equiv \mathrm{Tr}\rho\ln\rho-
\mathrm{Tr}\rho\ln\rho^{\prime}$ is the quantum relative entropy between two states $\rho$ and $\rho^{\prime}$.

We assume both the reservoir and the system are initially prepared in the thermal states, namely, $\rho_{R}=e^{-\beta_{R} \hat{H}_{R}}/Z_{R}$ for the reservoir and
$\rho_{S}=e^{-\beta_{S} \hat{H}_{S}}/Z_{S}$ for the system, with $\beta_{R}=1/k_B T_R$ and $\beta_S=1/k_B T_S$ the inverse temperatures
($k_B$ being the Boltzmann constant), and $Z_{R}$ and $Z_{S}$ the partition functions. In the process of evolution, though the system-environment correlation can be established between $S$ and $R_{n}$ before their interaction, the reduced state $\rho_{R_{n}}$ maintains the form of thermal state, i.e., $\rho_{R_{n}}=e^{-\beta_{R_n} H_{R}}/\text{Tr}[e^{-\beta_{R_n} H_{R}}]$ with the $n$-dependent inverse temperatures $\beta_{R_n}$.
In this case, we have $\mathrm{Tr}_{R_{n}}[(\rho_{R_{n}}^{\prime}-\rho_{R_{n}})\ln\rho_{R_{n}}]=\beta_{R_n}\Delta Q_{n}$ with
$\Delta Q_{n}=\mathrm{Tr}_{R_{n}}[(\rho_{R_{n}}-\rho_{R_{n}}^{\prime})H_{R}]$ denoting the heat flowing from reservoir element $R_{n}$
to the system $S$. Therefore, Eq. (\ref{DS}) can be written as
\begin{equation}\label{theorem}
\Delta S_{n}=D\left(\rho_{SR_{n}}^{\prime}\parallel\rho_{S_{n}}^{\prime}\rho_{R_{n}}\right)+\beta_{R_n}\Delta Q_{n}-I(\rho_{SR_{n}}).
\end{equation}
Without the creation of system-environment correlations during the dynamical process, for instance in the Markovian regime with no reservoir intracollisions, the mutual information $I(\rho_{SR_{n}})=0$ so that $\Delta S_{n}=D\left(\rho_{SR_{n}}^{\prime}\parallel\rho_{S_{n}}^{\prime}\rho_{R_{n}}\right)+\beta\Delta Q_{n}$,
with the two terms in the RHS being identified as the irreversible and reversible contributions
to the system entropy change due to heat exchanges \cite{enpro}.
In the non-Markovian regime, however, the system-environment correlations in terms of the mutual information $I(\rho_{SR_{n}})$ provide an additional contribute to the change of system entropy.

We notice that, at variance with the case of initial system-environment correlations \cite{jiangLandauer}, here we consider the problem of system entropy change from a different perspective, that is from the microscopic mechanism of dynamical memory effects. In fact, the term $I(\rho_{SR_{n}})$ in Eqs. (\ref{DS}) and (\ref{theorem}) characterizes the system-environment correlations established within the dynamical process, whose amount is determined by the intra-environment collision strength and is thus a signature of non-Markovianity of thermal reservoir \cite{ManXiaLo}. Therefore, instead of resorting to specific non-Markovianity quantifiers \cite{measure1,measure2,measure3,measure4,measure5,measure6}, we employ the established system-environment correlations as a figure of merit to reveal the quantitative condition for the validity of Landauer principle in the non-Markovian process.
To this purpose, from Eq. (\ref{theorem}), we obtain
\begin{equation}\label{theorem2}
\beta_{R_n}\Delta \widetilde{Q}_{n}=\Delta \widetilde{S}_{n}+D\left(\rho_{SR_{n}}^{\prime}\parallel\rho_{S_{n}}^{\prime}\rho_{R_{n}}\right)-I(\rho_{SR_{n}}),
\end{equation}
where $\Delta \widetilde{Q}_{n}=-\Delta Q_{n}$ is the heat dissipated to the reservoir and $\Delta \widetilde{S}_{n}=-\Delta S_{n}$ denotes the entropy decrease of the system. We refer to the equality of Eq. (\ref{theorem2}) as Landauer-like principle in the non-Markovian regime. From Eq. (\ref{theorem2}), we can immediately see that, when $I(\rho_{SR_{n}})\leq D\left(\rho_{SR_{n}}^{\prime}\parallel\rho_{S_{n}}^{\prime}\rho_{R_{n}}\right)$, the usual Landauer principle holds, that is $\beta_{R_n}\Delta \widetilde{Q}_{n}\geq\Delta \widetilde{S}_{n}$. In other words, the Landauer principle can still hold as long as the established system-environment correlations are smaller than such an upper bound. On the other hand, from Eq. (\ref{theorem2}) one finds that the condition
\begin{equation}\label{condi}
I(\rho_{SR_{n}})> D\left(\rho_{SR_{n}}^{\prime}\parallel\rho_{S_{n}}^{\prime}\rho_{R_{n}}\right),
\end{equation}
enables the violation of the Landauer principle during the dynamical process.

\subsection{Verification via a physical model}
In the following, using the collision model depicted in Fig. \ref{M1}, we verify both the equality of Eq. (\ref{theorem2}) [or, equivalently, Eq. (\ref{theorem})] and the condition for the violation of the Landauer principle given in Eq. (\ref{condi}).

Among the possible choices for the interaction between $S$ and a generic reservoir qubit $R_{n}$, we focus on a Heisenberg-like coherent interaction described by the Hamiltonian
\begin{equation}\label{H}
\hat{H}_\mathrm{int}=g(\hat{\sigma}_{x}^{S}\otimes\hat{\sigma}_{x}^{R_{n}}+\hat{\sigma}_{y}^{S}\otimes\hat{\sigma}_{y}^{R_{n}}+\hat{\sigma}_{z}^{S}\otimes\hat{\sigma}_{z}^{R_{n}}),
\end{equation}
where $\hat{\sigma}_{j}^{\mu}$ ($j=x,y,z$) is the Pauli operator, $g$ denotes a coupling constant and each collision is described by the unitary operator $\hat{U}_{SR_{n}}=e^{-i\hat{H}_\mathrm{int}\tau}$, $\tau$ being the collision time.
By means of the equality
\begin{equation}\label{eq}
e^{i\frac{\phi}{2}(\hat{\sigma}_{x}\otimes\hat{\sigma}_{x}+\hat{\sigma}_{y}\otimes\hat{\sigma}_{y}+\hat{\sigma}_{z}\otimes\hat{\sigma}_{z})}
=e^{-i\frac{\phi}{2}}(\cos\phi \ \hat{\mathbb{I}} +i\sin\phi \ \hat{\mathcal{S}}),
\end{equation}
where $\hat{\mathbb{I}}$ is the identity operator and $\hat{\mathcal{S}}$ the two-qubit swap operator such that $\hat{\mathcal{S}}\left|\psi_{1}\right\rangle\otimes\left|\psi_{2}\right\rangle=\left|\psi_{2}\right\rangle\otimes\left|\psi_{1}\right\rangle$
for all $\left|\psi_{1}\right\rangle,\left|\psi_{2}\right\rangle\in \mathbb{C}^{2}$, the unitary time evolution operator can be written as
\begin{equation}\label{swapSR1}
\hat{U}_{SR_{n}}=(\cos J)\ \hat{\mathbb{I}}_{SR_{n}}+i(\sin J)\ \hat{\mathcal{S}}_{SR_{n}},
\end{equation}
where $J=2g\tau$ is a dimensionless interaction strength between $S$ and $R_{n}$ which is supposed to be the same for any $n=1,2,\ldots,N$.
It is immediate to see that $J=\pi/2$ induces a complete swap between the state of $S$ and that of $R_{n}$. Thus, $0<J<\pi/2$ means a partial swap conveying the intuitive idea that, at each collision, part of the information contained in the state of $S$ is transferred into $R_{n}$.
In the ordered basis $\mathcal{B}_{SR_{n}}=\{\left|00\right\rangle_{SR_{n}},\left|01\right\rangle_{SR_{n}},\left|10\right\rangle_{SR_{n}},\left|11\right\rangle_{SR_{n}}\}$, $\hat{U}_{SR_{n}}$ reads
\begin{equation}\label{swapSR2}
\hat{U}_{SR_{n}}=\left(
                   \begin{array}{cccc}
                     e^{iJ} & 0 & 0 & 0 \\
                     0 & \cos J & i\sin J & 0 \\
                     0 & i\sin J & \cos J & 0 \\
                     0 & 0 & 0 & e^{iJ} \\
                   \end{array}
                 \right).
\end{equation}
The interaction between two nearest-neighbor reservoir qubits $R_{n}$ and $R_{n+1}$ is also described by an operation similar to that of Eq.~(\ref{swapSR2}), namely
\begin{equation}\label{swapRR}
\hat{V}_{R_{n}R_{n+1}}=\left(
                   \begin{array}{cccc}
                     e^{i\Omega} & 0 & 0 & 0 \\
                     0 & \cos \Omega & i\sin \Omega & 0 \\
                     0 & i\sin \Omega & \cos \Omega & 0 \\
                     0 & 0 & 0 & e^{i\Omega} \\
                   \end{array}
                 \right),
\end{equation}
where $0\leq\Omega\leq\pi/2$ is the dimensionless $R_{n}$-$R_{n+1}$ interaction strength being independent of $n$.

\begin{figure}[tbp]
\begin{center}
{\includegraphics[width=0.43\textwidth]{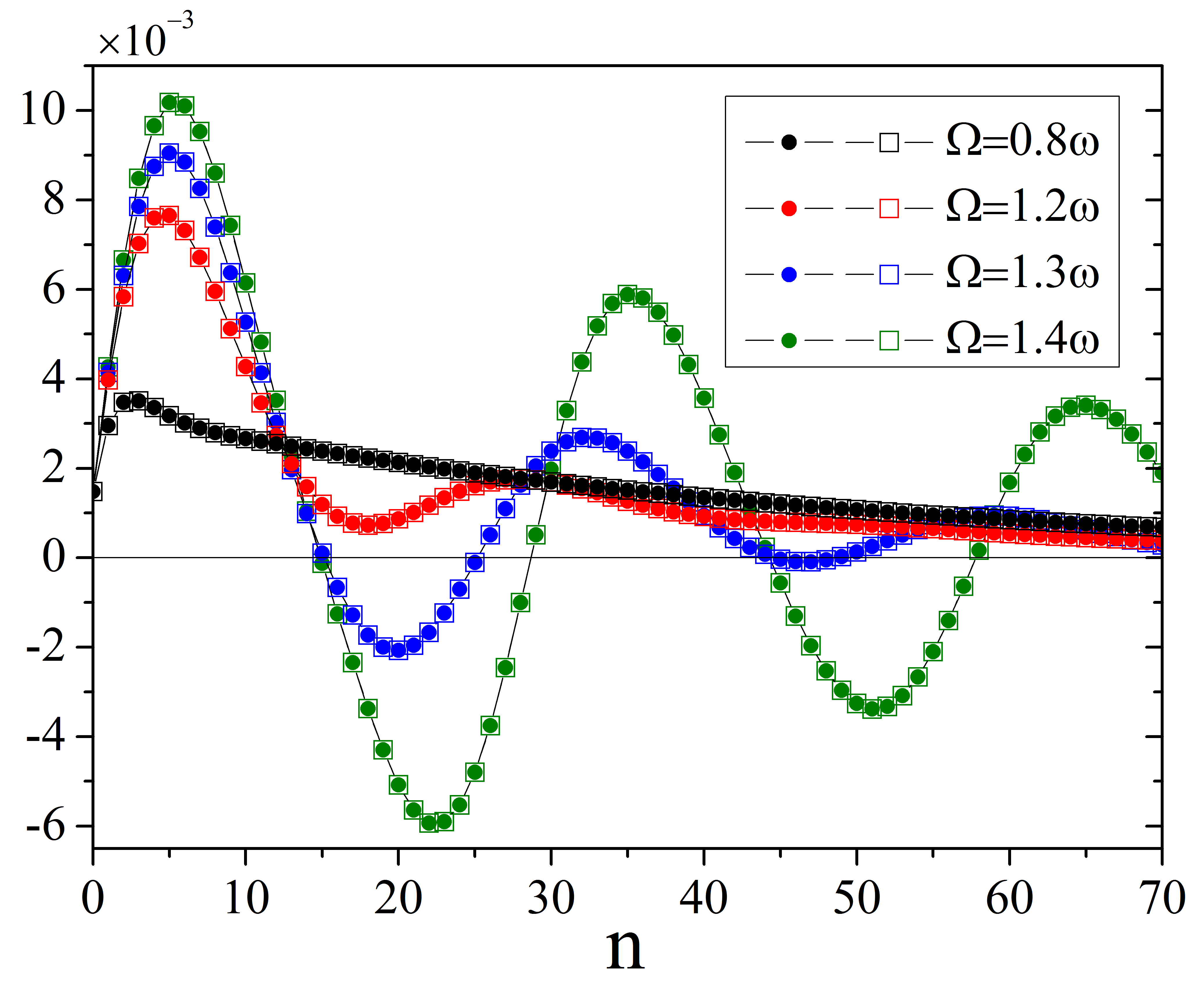}}
\end{center}
\caption{(Color online) The LHS (circle) and RHS (square) of Eq. (\ref{theorem2}) against the collision number $n$ for different inter-environment collision strength $\Omega$.
The initial temperatures of the system and the reservoir are chosen as $T_{S}=3\omega$ and $T_{R}=\omega$, respectively. The coupling constant between the system and the reservoir qubits is set as $J=0.1\omega$.}
\label{equality}
\end{figure}

After deriving $\rho_{S_{n}}$, $\rho_{R_{n}}$, $\rho_{S_{n}}^{\prime}$ and $\rho_{R_{n}}^{\prime}$,
we can obtain all the terms appearing in the equality of Eq. (\ref{theorem2}) and are thus in the position to explore their dynamics with respect to the collision number $n$.
The equality of Eq. (\ref{theorem2}) is verified in Fig. \ref{equality} by comparing its LHS and RHS in the dynamical process, which exhibit complete coincidence. Therefore, this equality supplies a reliable information measure for the heat dissipation of an information erasure process in a non-Markovian environment.

\begin{figure}[tbp]
\begin{center}
{\includegraphics[width=0.43\textwidth]{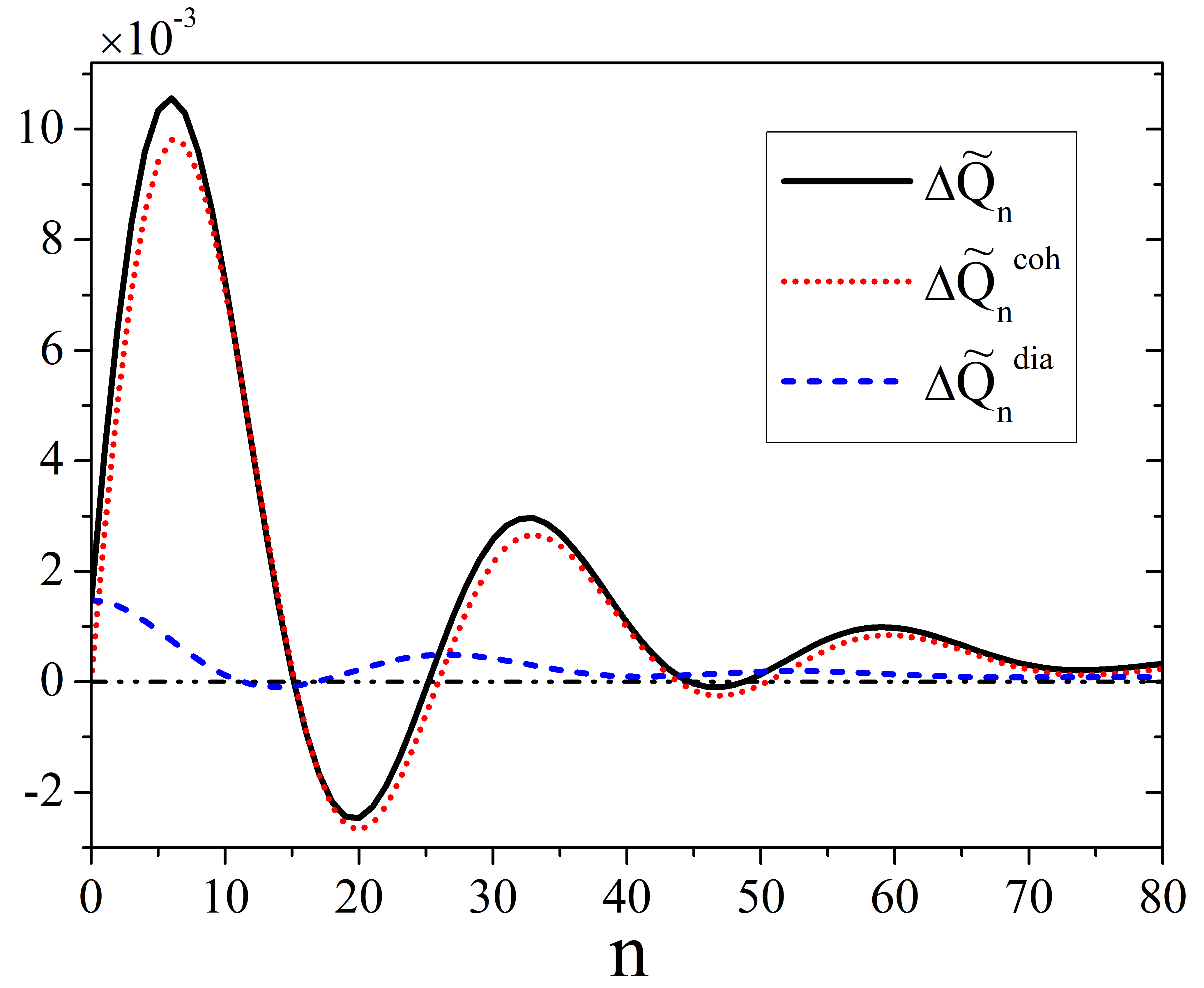}}
\end{center}
\caption{(Color online) The heat $\Delta \widetilde{Q}_{n}$ absorbed by the reservoir and its two contributions $\Delta \widetilde{Q}_{n}^{dia}$ and $\Delta \widetilde{Q}_{n}^{coh}$ against the number of collisions $n$ for $\Omega=1.3\omega$. Other parameters are the same as those of Fig. \ref{equality}.}
\label{nega}
\end{figure}

\begin{figure}[tbp]
\begin{center}
{\includegraphics[width=0.44\textwidth]{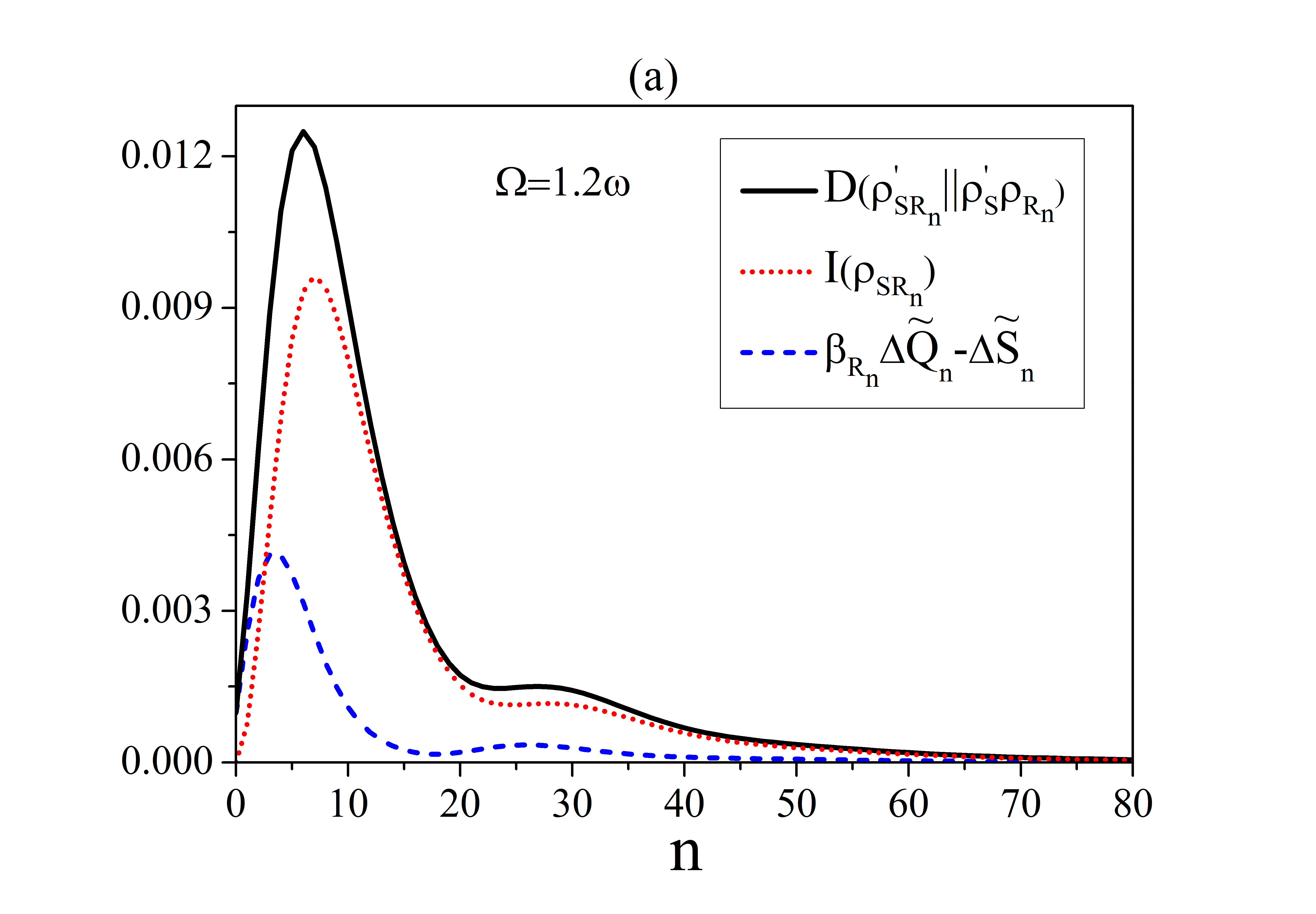}}
{\includegraphics[width=0.44\textwidth]{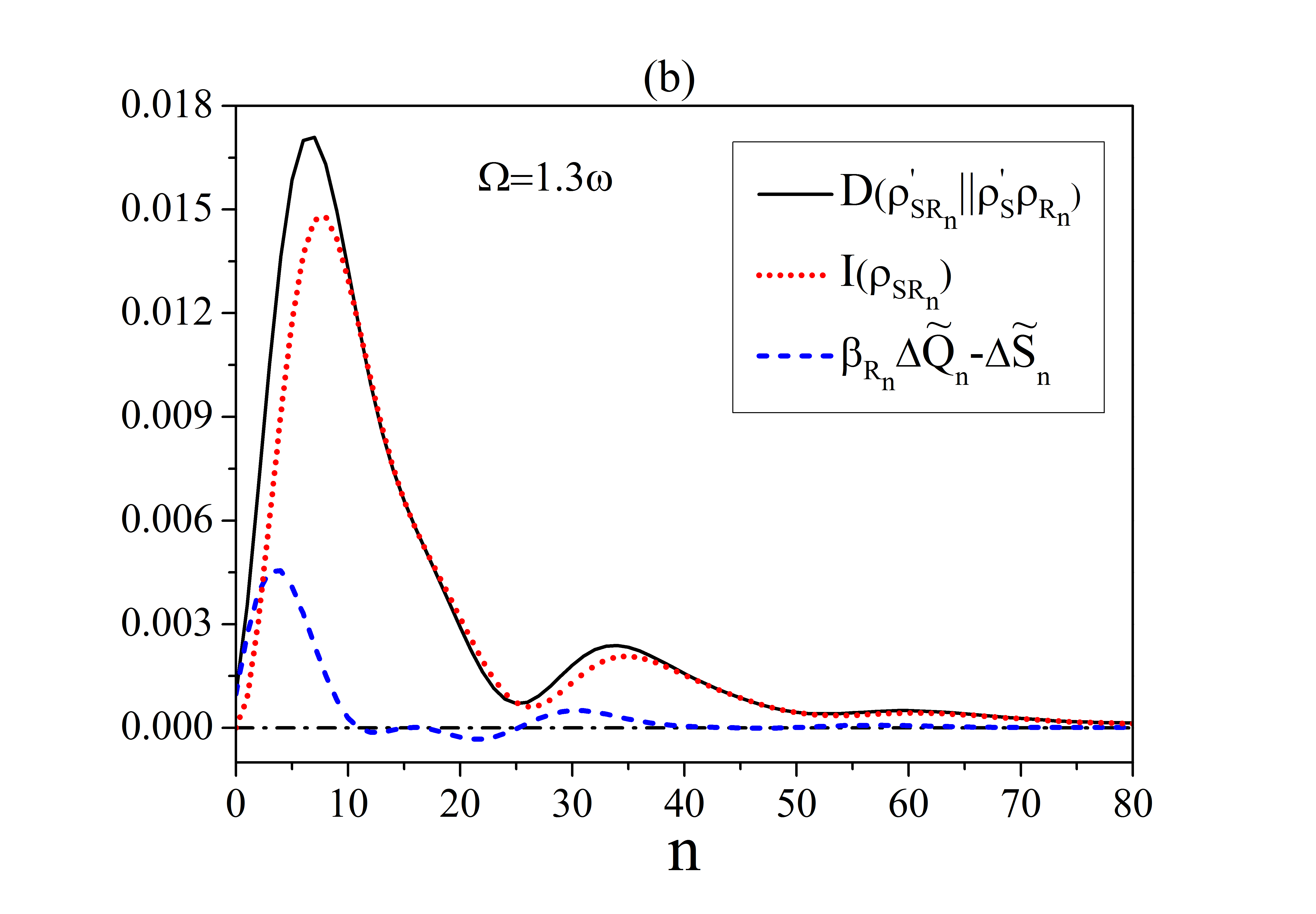}}
{\includegraphics[width=0.44\textwidth]{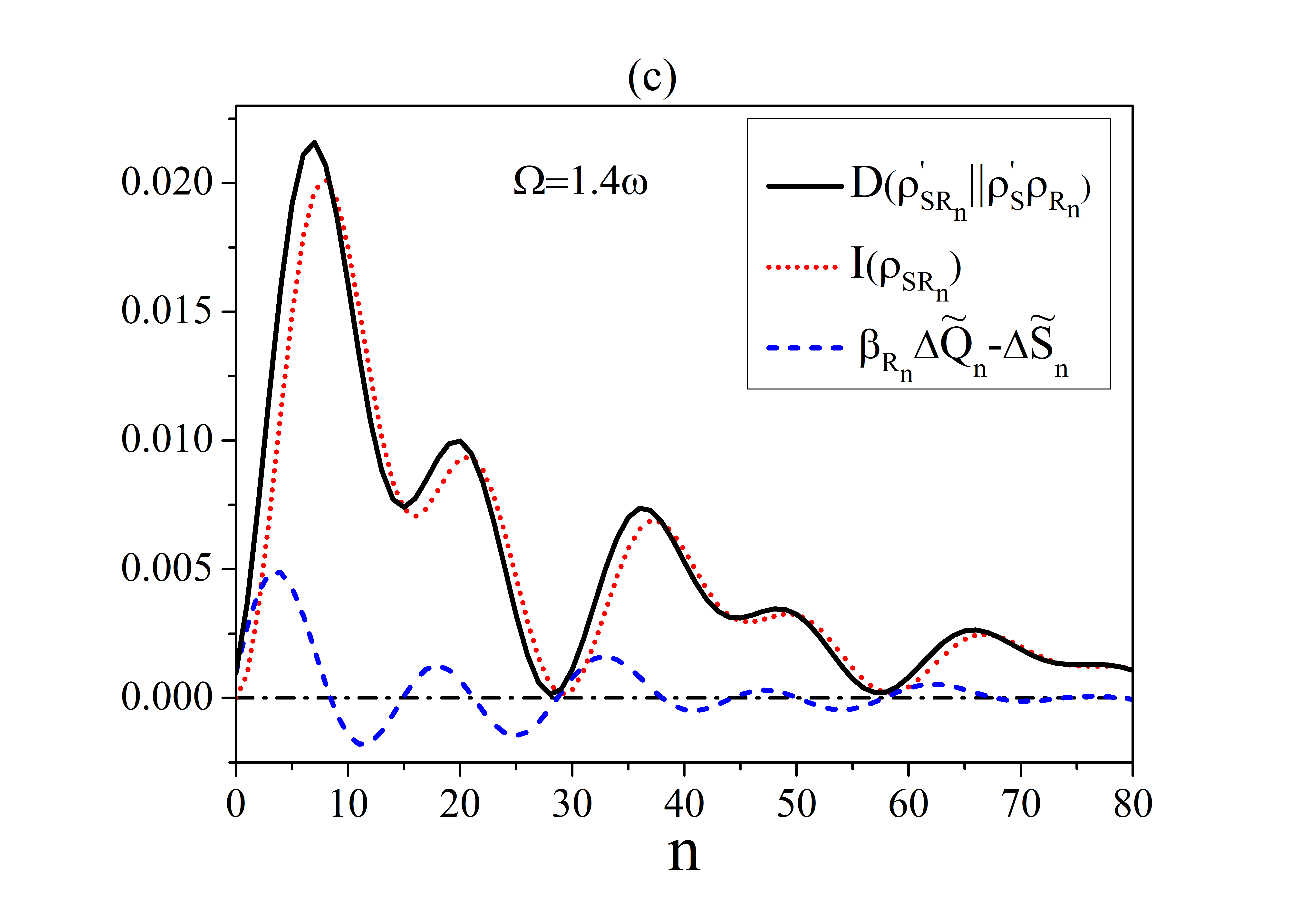}}
\end{center}
\caption{(Color online) The established system-environment correlation in terms of $I(\rho_{SR_{n}})$ , the entropy production
$D\left(\rho_{SR_{n}}^{\prime}\|\rho_{S}^{\prime}\rho_{R_{n}}\right)$, and the difference of $\beta_{R_n}\Delta \widetilde{Q}_{n}-\Delta \widetilde{S}_{n}$ against the collision number $n$ for different inter-environment collision strength $\Omega$.
The other parameters are the same as that given in Fig. \ref{equality}. When $I(\rho_{SR_{n}})>D\left(\rho_{SR_{n}}^{\prime}\|\rho_{S}^{\prime}\rho_{R_{n}}\right)$, we have $\beta_{R_n}\Delta \widetilde{Q}_{n}<\Delta \widetilde{S}_{n}$, i.e. the Landauer principle is violated.}
\label{viol}
\end{figure}

In Fig. \ref{equality}, though the initial temperature $T_{S}$ of the system is set larger than $T_{R}$ of the reservoir, we note that $\Delta \widetilde{Q}_{n}$ can attain negative values indicating the backflow of heat from the reservoir to the system in the evolution for relatively large values of $\Omega$ (e.g., $\Omega=1.3\omega$, $1.4\omega$). Actually, $\Delta \widetilde{Q}_{n}$ can be divided into two different contributions
\begin{equation}\label{heatR}
\Delta \widetilde{Q}_{n}=\Delta \widetilde{Q}_{n}^{dia}+\Delta \widetilde{Q}_{n}^{coh},
\end{equation}
where
\begin{equation}\label{heatRelem}
\Delta \widetilde{Q}_{n}^{dia}=\omega\sin^{2}(J)(\rho_{33}-\rho_{22}),\quad
\Delta \widetilde{Q}_{n}^{coh}=\omega\mathrm{Im}(\rho_{23})\sin(2J),
\end{equation}
are the heats determined, respectively, by the diagonal and coherent (off-diagonal) elements of the state $\rho_{SR_{n}}$ of $S$-$R_{n}$ before their collisions.
In Eq. (\ref{heatRelem}), we have used the ordered basis $\mathcal{B}_{SR_{n}}$ of $S$-$R_{n}$ by setting $\left\{\left|\overline{1}\right\rangle=\left|00\right\rangle,\left|\overline{2}\right\rangle=\left|01\right\rangle,
\left|\overline{3}\right\rangle=\left|10\right\rangle,\left|\overline{4}\right\rangle=\left|11\right\rangle\right\}$ and
$\rho_{kl}=\langle \overline{k}|\rho_{SR_{n}}|\overline{l}\rangle$ with $k, l=1, 2, 3, 4$.
The nonzero coherent term $\rho_{23}$ of $\rho_{SR_{n}}$ is a direct witness of the correlations between $S$ and $R_{n}$, which in turn gives the correlation-dependent heat $\Delta \widetilde{Q}_{n}^{coh}$. For relatively small values of $\Omega$, the established system-environment correlations are weak, so the contribution $\Delta \widetilde{Q}_{n}^{dia}$ plays a major role in determining the behavior of the total heat $\Delta \widetilde{Q}_{n}$.
Differently, when $\Omega$ is sufficiently large the behavior of $\Delta \widetilde{Q}_{n}$, especially its transition from positive to negative values, is mainly determined by the contribution $\Delta \widetilde{Q}_{n}^{coh}$, as highlighted in Fig. \ref{nega}.

Now we check the condition for the violation of the Landauer principle given in Eq. (\ref{condi}) by comparing the values of $I(\rho_{SR_{n}})$ and $D\left(\rho_{SR_{n}}^{\prime}\|\rho_{S}^{\prime}\rho_{R_{n}}\right)$ for different intra-reservoir collision strengths $\Omega$.
As shown in Fig. \ref{viol}(a), for relatively small values of $\Omega$ (e.g., $\Omega=1.2\omega$), $I(\rho_{SR_{n}})$ is smaller than $D\left(\rho_{SR_{n}}^{\prime}\|\rho_{S}^{\prime}\rho_{R_{n}}\right)$ in the whole dynamical process. This means that $\beta_{R_n}\Delta \widetilde{Q}_{n}$ remains larger than $\Delta \widetilde{S}_{n}$, implying that the Landauer principle still holds although there exist nonzero established system-environment correlations at later times (memory effects).
By contrast, as shown in Fig. \ref{viol}(b) and (c), relatively large values of $\Omega$ (e.g., $\Omega=1.3\omega$, $1.4\omega$) can make $I(\rho_{SR_{n}})$ supersede $D\left(\rho_{SR_{n}}^{\prime}\|\rho_{S}^{\prime}\rho_{R_{n}}\right)$ during some time intervals of the evolution when the Landauer principle is thus violated, being $\beta_{R_n}\Delta \widetilde{Q}_{n}<\Delta \widetilde{S}_{n}$.

\section{Landauer principle in multiple non-Markovian reservoirs} \label{sec3}

\begin{figure}[tbp]
\begin{center}
{\includegraphics[width=0.47\textwidth]{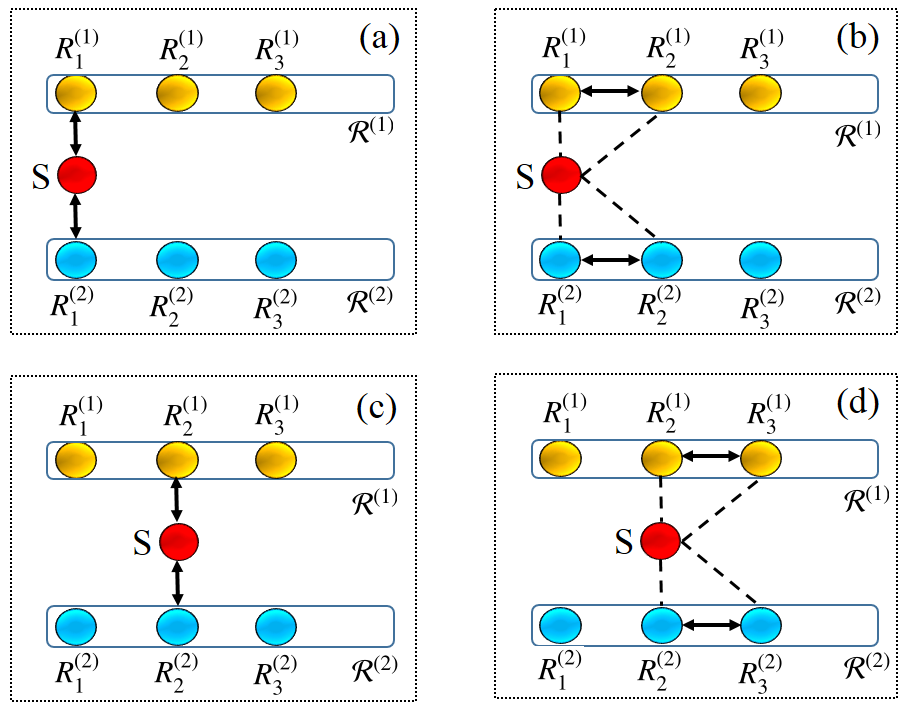}}
\end{center}
\caption{(Color online) (a) System $S$ collides in a sequence with qubits $R_{1}^{(1)}$, $R_{1}^{(2)}$ in reservoirs $\mathcal{R}^{(1)}$, $\mathcal{R}^{(2)}$, respectively, so that they get correlated. The established correlations are denoted by dashed lines. (b) The intracollisions of $R_{1}^{(1)}$-$R_{2}^{(1)}$ and $R_{1}^{(2)}$-$R_{2}^{(2)}$ take place, generating correlations among $S$, $R_{1}^{(1)}$, $R_{1}^{(2)}$, $R_{2}^{(1)}$ and $R_{2}^{(2)}$. (c) Reservoir qubits $R_{1}^{(1)}$ and $R_{1}^{(2)}$ are traced out, while the correlations of $S$-$R_{2}^{(1)}$-$R_{2}^{(2)}$ remain. Then, $S$ collides with both $R_{2}^{(1)}$ and $R_{2}^{(2)}$ which is followed by intracollisions of $R_{2}^{(1)}$-$R_{3}^{(1)}$, $R_{2}^{(2)}$-$R_{3}^{(2)}$ in panel (d).}
\label{M2}
\end{figure}

The previous results obtained for a single non-Markovian reservoir leaves now open the question whether they remain qualitatively the same for a composite environment of multiple reservoirs, the only difference emerging from the quantitative side. In this section we investigate this aspect.
In the following, we generalize the results obtained in a single reservoir to multiple reservoirs, where the quantum system $S$ is coupled to $M$ finite-size heat reservoirs, i.e., $\mathcal{R}^{(1)},$ $\mathcal{R}^{(2)},$...$\mathcal{R}^{(M)}$,
with each one consists of $N$ identical qubits $R_{1}^{(m)},R_{2}^{(m)},\ldots,R_{N}^{(m)}$ ($m=1,2,3...M$). The system and generic reservoir qubit are still described by the Hamiltonians given in Eq. (\ref{HSR}). The model with $M=2$ is illustrated in Fig.~\ref{M2} for the first two steps of collisions.

To begin with, the system $S$ collides with the first qubits $R_{1}^{(1)},$ $R_{1}^{(2)},$...$R_{1}^{(M)}$ in all the $M$ reservoirs and then the intra-reservoir collisions of $R_{1}^{(1)}$-$R_{2}^{(1)}$, $R_{1}^{(2)}$-$R_{2}^{(2)}$, $\dots$, $R_{1}^{(M)}$-$R_{2}^{(M)}$ occurs leading to the constructions of correlations among the various parts $S$, $R_{2}^{(1)}$, $R_{2}^{(2)}$, $\ldots$, $R_{2}^{(M)}$ prior to their collisions.
Generally, the state $\rho_{SR_{n}^{(1)}R_{n}^{(2)}...R_{n}^{(M)}}$ of $S,$ $R_{n}^{(1)},$ $R_{n}^{(2)},$...$R_{n}^{(M)}$ are correlated before their collisions for $n\geq2$ and such correlations are characterized by the mutual information $I(\rho_{SR_{n}^{(1)}R_{n}^{(2)}\ldots R_{n}^{(M)}})=S(\rho_{S_n})+\sum_{m=1}^{M}S(\rho_{R_{n}^{(m)}})-S(\rho_{SR_{n}^{(1)}R_{n}^{(2)}\ldots R_{n}^{(M)}})$ with
$\rho_{S_{n}}$ and $\rho_{R_{n}^{(m)}}$ the marginal states of $S$ and $R_{n}^{(m)}$.
Governed by the unitary actions $\hat{U}_{SR_{n}^{(1)}}$, $\hat{U}_{SR_{n}^{(2)}}$, $...$, $\hat{U}_{SR_{n}^{(M)}}$ for the interactions of $S$-$R_{n}^{(1)}$, $S$-$R_{n}^{(2)}$, $\ldots$ $S$-$R_{n}^{(M)}$, the state $\rho_{SR_{n}^{(1)}R_{n}^{(2)}\ldots R_{n}^{(M)}}$ is transformed to $\rho_{SR_{n}^{(1)}R_{n}^{(2)}\ldots R_{n}^{(M)}}^{\prime}=\hat{U}_{SR_{n}^{(M)}}\ldots \hat{U}_{SR_{n}^{(2)}}\hat{U}_{SR_{n}^{(1)}}\rho_{SR_{n}^{(1)}R_{n}^{(2)}\ldots R_{n}^{(M)}}\hat{U}_{SR_{n}^{(1)}}^{\dag}\hat{U}_{SR_{n}^{(2)}}^{\dag}\ldots\hat{U}_{SR_{n}^{(M)}}^{\dag}$.
The marginal states of $S$ and $R_{n}^{(m)}$ after the collisions are labeled as $\rho_{S_{n}}^{\prime}=\mathrm{Tr}_{R_{n}^{(1)}R_{n}^{(2)}\ldots R_{n}^{(M)}}\rho_{SR_{n}^{(1)}R_{n}^{(2)}\ldots R_{n}^{(M)}}^{\prime}$ and $\rho_{R_{n}^{(m)}}^{\prime}=\mathrm{Tr}_{S\overline{R}_{n}^{(m)}}\rho_{SR_{n}^{(1)}R_{n}^{(2)}\ldots R_{n}^{(M)}}^{\prime}$ with $\overline{R}_{n}^{(m)}$
the reservoir qubits other than $R_{n}^{(m)}$.
By the fact that the total von Neumann entropy of the system and reservoirs under the unitary actions retains invariant, namely, $S(\rho_{SR_{n}^{(1)}R_{n}^{(2)}...R_{n}^{(M)}})=S(\rho_{SR_{n}^{(1)}R_{n}^{(2)}...R_{n}^{(M)}}^{\prime})$,
we can derive the entropy change $\Delta S_{n}$ of the system as (see Appendix \ref{appenB} for details)
\begin{eqnarray} \label{multiDS}
  \Delta S_{n} &=& S\left(\rho_{S_{n}}^{\prime}\right)-S\left(\rho_{S_{n}}\right) \nonumber\\
   &=&  D\left(\rho_{SR_{n}^{(1)}R_{n}^{(2)}\ldots R_{n}^{(M)}}^{\prime}\parallel\rho_{S_{n}}^{\prime}\prod_{m=1}^{M}\rho_{R_{n}^{(m)}}\right)\\
&+&  \sum_{m=1}^{M}\mathrm{Tr}_{R_{n}^{(m)}}\left(\rho_{R_{n}^{(m)}}^{\prime}-\rho_{R_{n}^{(m)}}\right)
\ln\rho_{R_{n}^{(m)}}-I\left(\rho_{SR_{n}^{(1)}R_{n}^{(2)}\ldots R_{n}^{(M)}}\right),\nonumber
\end{eqnarray}
where $D\left(\rho_{SR_{n}^{(1)}R_{n}^{(2)}...R_{n}^{(M)}}^{\prime}\parallel\rho_{S_{n}}^{\prime}\prod_{m=1}^{M}\rho_{R_{n}^{(m)}}\right)$ is the irreversible entropy production of the system \cite{enpro}. Here, we still assume both the system and reservoirs are initially prepared in the thermal states but the reservoirs may have different temperatures, namely, for the $m$-th reservoir its state $\rho_{R^{(m)}}=e^{-\beta^{(m)} \hat{H}_{R}}/\text{Tr}[e^{-\beta^{(m)} \hat{H}_{R}}]$
with $\beta^{(m)}=1/k_B T^{(m)}$ the inverse temperature. In the process of evolution,
though the system-reservoir correlations can be established between $S$ and $R_{n}^{(m)}$ before their interactions, the reduced states $\rho_{R_{n}^{(m)}}$ still retain
the forms of thermal states, i.e., $\rho_{R_{n}^{(m)}}=e^{-\beta^{(m)}_{n} \hat{H}_{R}}/\text{Tr}[e^{-\beta^{(m)}_{n} \hat{H}_{R}}]$
with the $n$-dependent inverse temperatures $\beta^{(m)}_{n}$.
In this case, we have $\mathrm{Tr}_{R_{n}^{(m)}}\left(\rho_{R_{n}^{(m)}}^{\prime}-\rho_{R_{n}^{(m)}}\right)
\ln\rho_{R_{n}^{(m)}}=\beta_{n}^{(m)}\Delta Q_{n}^{(m)}$ with
$\Delta Q_{n}^{(m)}=\mathrm{Tr}_{R_{n}^{(m)}}\left[\left(\rho_{R_{n}^{(m)}}-\rho_{R_{n}^{(m)}}^{\prime}\right)\hat{H}_{R}\right]$ denoting the heat flowing from reservoir qubit $R_{n}^{(m)}$ to the system $S$. Therefore, the Eq.(\ref{multiDS}) is reformulated as
\begin{eqnarray}\label{multitheorem}
\Delta S_{n}&=&D\left(\rho_{SR_{n}^{(1)}R_{n}^{(2)}...R_{n}^{(M)}}^{\prime}\parallel\rho_{S_{n}}^{\prime}\prod_{m=1}^{M}\rho_{R_{n}^{(m)}}\right)
  +\sum_{m=1}^{M}\beta_{n}^{(m)}\Delta Q_{n}^{(m)}\nonumber\\
  &&-I\left(\rho_{SR_{n}^{(1)}R_{n}^{(2)}...R_{n}^{(M)}}\right).
\end{eqnarray}
Without the construction of system-reservoir correlations in the dynamical process, for instance, in the Markovian regime with $\Omega=0$, the mutual information $I\left(\rho_{SR_{n}^{(1)}R_{n}^{(2)}...R_{n}^{(M)}}\right)=0$ so that we have $\Delta S_{n}=D\left(\rho_{SR_{n}^{(1)}R_{n}^{(2)}...R_{n}^{(M)}}^{\prime}\parallel\rho_{S_{n}}^{\prime}\prod_{m=1}^{M}\rho_{R_{n}^{(m)}}\right)
  +\sum_{m=1}^{M}\beta_{n}^{(m)}\Delta Q_{n}^{(m)}$
with the first term and the other $M$ terms in RHS being identified as the irreversible and reversible contributions, respectively,
to the system entropy change due to heat exchanges \cite{enpro}.
In the non-Markovian regime, however, the system entropy change will also be contributed by the system-reservoir correlations
in terms of the mutual information $I\left(\rho_{SR_{n}^{(1)}R_{n}^{(2)}...R_{n}^{(M)}}\right)$.

\begin{figure}[t!]
\begin{center}
{\includegraphics[width=0.45 \textwidth]{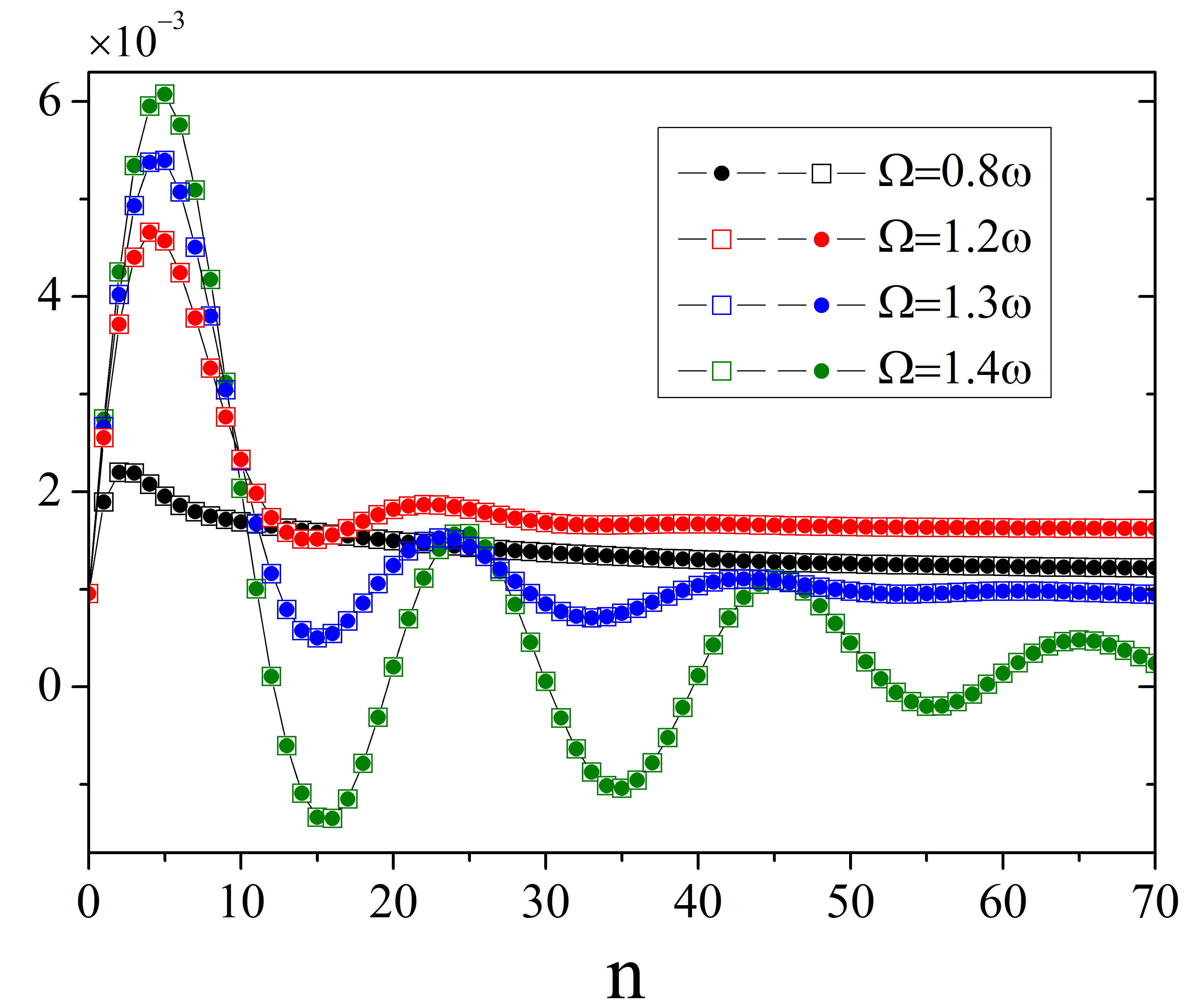}}
\end{center}
\caption{(Color online) The LHS (circle) and RHS (square) of Eq. (\ref{multitheorem2}) against the collision number $n$. The initial temperatures of the system and the two reservoirs are chosen as $T_{S}=2\omega$, $T^{(1)}=3\omega$ and $T^{(2)}=\omega$, respectively. Here, we assume the same strengths $\Omega$ for the intracollisions of
$R_{n}^{(1)}$-$R_{n+1}^{(1)}$ and $R_{n}^{(2)}$-$R_{n+1}^{(2)}$ and the same coupling constants $J=0.1\omega$ for $S$-$R_{n}^{(1)}$ and $S$-$R_{n}^{(2)}$.}
\label{multiequality}
\end{figure}

We now connect the equality (\ref{multitheorem}) to the Landauer principle in the presence of multiple non-Markovian reservoirs.
From Eq. (\ref{multitheorem}), we obtain that
\begin{eqnarray}\label{multitheorem2}
\sum_{m=1}^{M}\beta_{n}^{(m)}\Delta \widetilde{Q}_{n}^{(m)}&=&\Delta \widetilde{S}_{n}+D\left(\rho_{SR_{n}^{(1)}R_{n}^{(2)}...R_{n}^{(M)}}^{\prime}\parallel\rho_{S_{n}}^{\prime}\prod_{m=1}^{M}\rho_{R_{n}^{(m)}}\right)\nonumber\\
&&-I\left(\rho_{SR_{n}^{(1)}R_{n}^{(2)}...R_{n}^{(M)}}\right),
\end{eqnarray}
where $\Delta \widetilde{Q}_{n}^{(m)}=-\Delta Q_{n}^{(m)}$ is the heat dissipated to $R_{n}^{(m)}$ and $\Delta \widetilde{S}_{n}=-\Delta S_{n}$ denotes
the entropy decrease of the system. We call the equality (\ref{multitheorem2}) Landauer-like principle in the multiple non-Markovian reservoirs.
From Eq. (\ref{multitheorem2}), we can immediately obtain that when $I\left(\rho_{SR_{n}^{(1)}R_{n}^{(2)}...R_{n}^{(M)}}\right)\leq D\left(\rho_{SR_{n}^{(1)}R_{n}^{(2)}...R_{n}^{(M)}}^{\prime}\parallel\rho_{S_{n}}^{\prime}\prod_{m=1}^{M}\rho_{R_{n}^{(m)}}\right)$, the Landauer principle still holds, namely, $\sum_{m=1}^{M}\beta_{n}^{(m)}\Delta \widetilde{Q}_{n}^{(m)}\geq\Delta \widetilde{S}_{n}$.
At the same time, the condition for the violation of the Landauer principle
is obtained as
\begin{equation}\label{cond-multi}
 I\left(\rho_{SR_{n}^{(1)}R_{n}^{(2)}...R_{n}^{(M)}}\right)> D\left(\rho_{SR_{n}^{(1)}R_{n}^{(2)}...R_{n}^{(M)}}^{\prime}\parallel\rho_{S_{n}}^{\prime}\prod_{m=1}^{M}\rho_{R_{n}^{(m)}}\right).
\end{equation}

\begin{figure}[b!]
\begin{center}
{\includegraphics[width=0.44\textwidth]{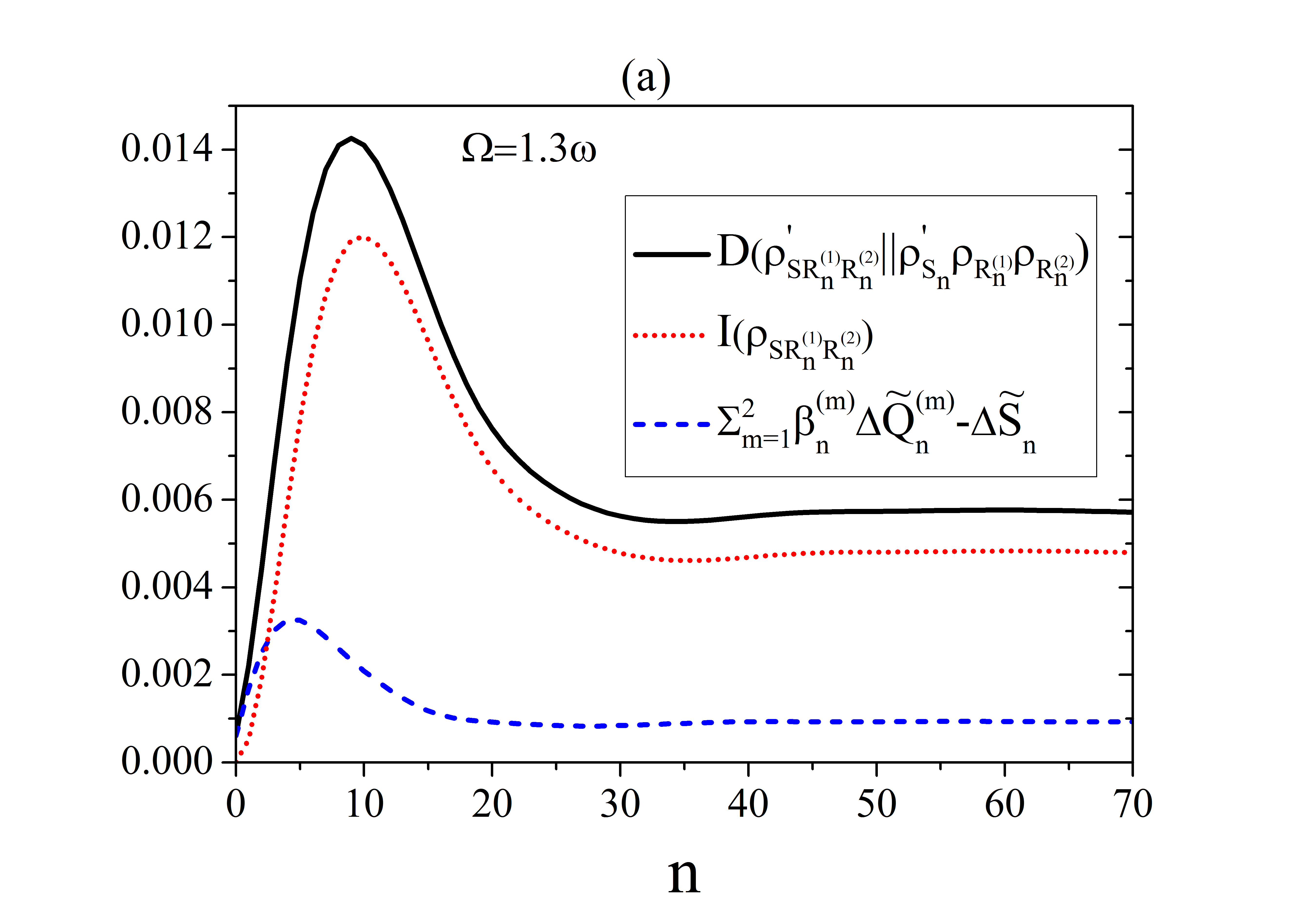}}
{\includegraphics[width=0.44\textwidth]{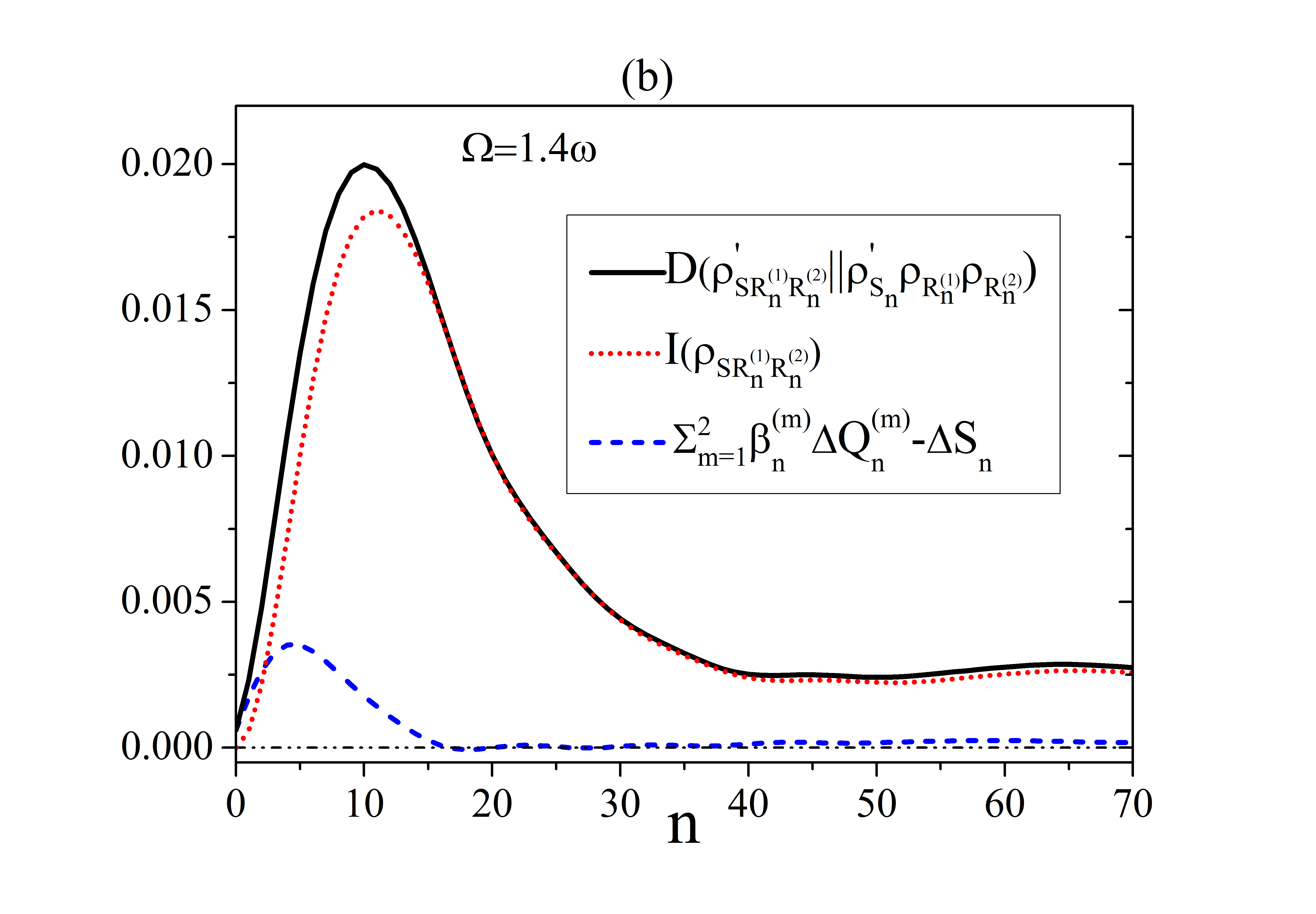}}
{\includegraphics[width=0.44\textwidth]{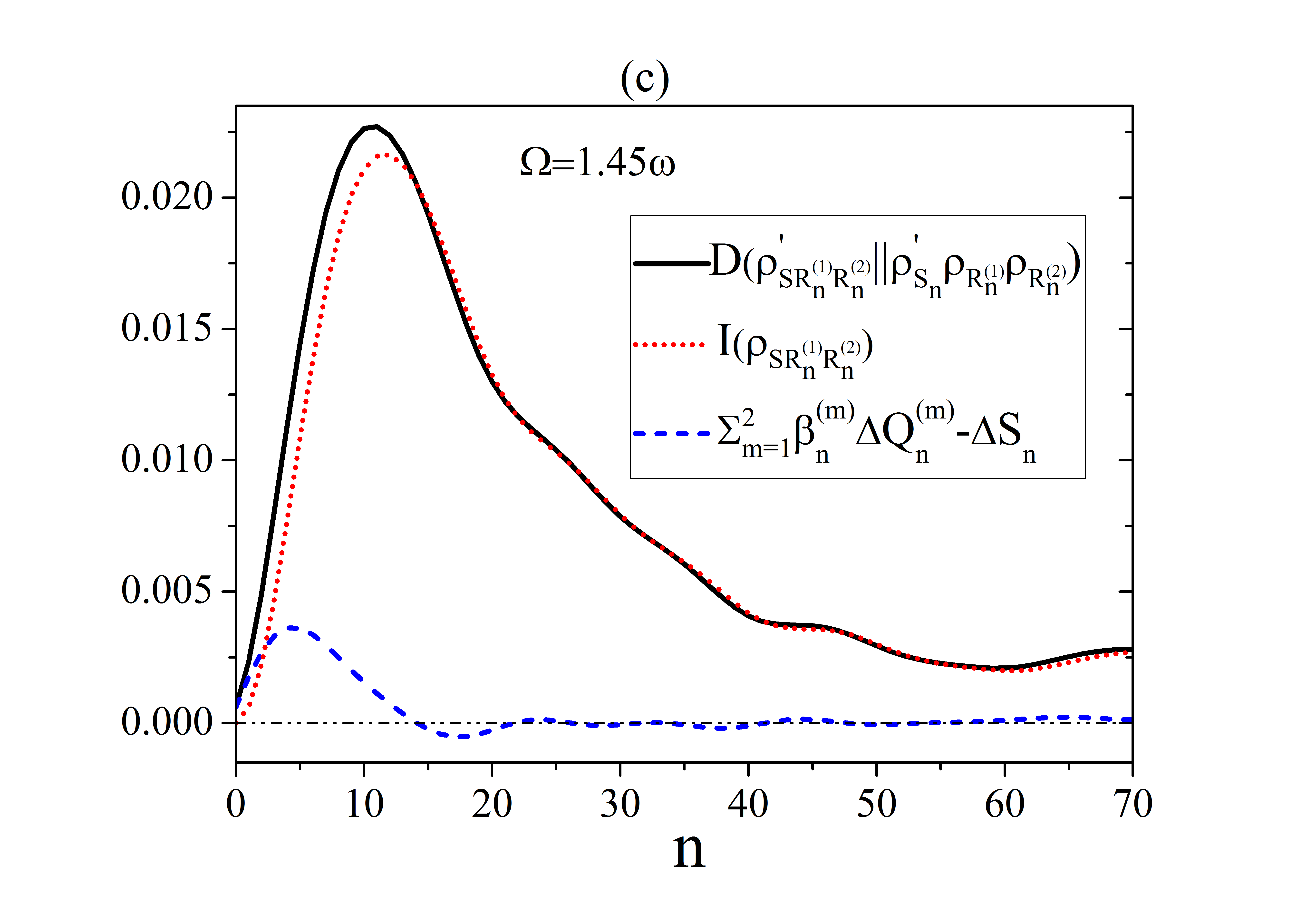}}
\end{center}
\caption{(Color online) Established system-environment correlations $I(\rho_{SR_{n}^{(1)}R_{n}^{(2)}})$, entropy production $D\left(\rho_{SR_{n}^{(1)}R_{n}^{(2)}}^{\prime}\parallel\rho_{S_{n}}^{\prime}\rho_{R_{n}^{(1)}}\rho_{R_{n}^{(2)}}\right)$, and the difference $\sum_{m=1}^{2}\beta_{n}^{(m)}\Delta \widetilde{Q}_{n}^{(m)}-\Delta \widetilde{S}_{n}$ as a function of the number of collisions $n$ for different inter-environment collision strengths $\Omega$. The other parameters are the same as those given in Fig. \ref{multiequality}.
When $I(\rho_{SR_{n}^{(1)}R_{n}^{(2)}})>D\left(\rho_{SR_{n}^{(1)}R_{n}^{(2)}}^{\prime}\parallel\rho_{S_{n}}^{\prime}\rho_{R_{n}^{(1)}}\rho_{R_{n}^{(2)}}\right)$, one has $\sum_{m=1}^{2}\beta_{n}^{(m)}\Delta \widetilde{Q}_{n}^{(m)}<\Delta \widetilde{S}_{n}$, i.e. Landauer principle violation.}
\label{multiviol}
\end{figure}

Based on the model with $M=2$ as illustrated in Fig. \ref{M2}, we verify the equality (\ref{multitheorem2}) and the condition (\ref{cond-multi}) for the violation of Landauer principle. For the interactions between $S$ and reservoir qubits as well as intra-interactions between reservoirs qubits, we still adopt the Heisenberg-like interaction described by the Hamiltonian (\ref{H}). Specifically, we apply $\hat{U}_{SR_{n}^{(1)}}=\hat{U}_{SR_{n}^{(2)}}\equiv\hat{U}_{SR_{n}}$ in Eq. (\ref{swapSR2}) for the interactions of $S$-$R_{n}^{(1)}$ and $S$-$R_{n}^{(2)}$, while $\hat{V}_{R_{n}^{(1)}R_{n+1}^{(1)}}=\hat{V}_{R_{n}^{(2)}R_{n+1}^{(2)}}\equiv\hat{V}_{R_{n}R_{n+1}}$ in Eq. (\ref{swapRR}) for the interactions of $R_{n}^{(1)}$-$R_{n+1}^{(1)}$ and $R_{n}^{(2)}$-$R_{n+1}^{(2)}$. A comparison between the LHS and RHS of the equality of Eq. (\ref{multitheorem2}) is shown in Fig. \ref{multiequality}, where one can see their complete coincidences implying the validity of equality of Eq. (\ref{multitheorem2}).
In Fig. \ref{multiviol}(a), (b) and (c), we check the condition of Eq. (\ref{cond-multi}) that leads to Landauer principle violation in the presence of two non-Markovian reservoirs by comparing the values of
$I(\rho_{SR_{n}^{(1)}R_{n}^{(2)}})$ and $D\left(\rho_{SR_{n}^{(1)}R_{n}^{(2)}}^{\prime}\parallel\rho_{S_{n}}^{\prime}\rho_{R_{n}^{(1)}}\rho_{R_{n}^{(2)}}\right)$ for different intracollision strengths $\Omega$.
As displayed in Fig. \ref{multiviol}(a), for relatively small values of $\Omega$ (e.g., $\Omega=1.2\omega$), the amount of $I(\rho_{SR_{n}^{(1)}R_{n}^{(2)}})$ is always less than $D\left(\rho_{SR_{n}^{(1)}R_{n}^{(2)}}^{\prime}\parallel\rho_{S_{n}}^{\prime}\rho_{R_{n}^{(1)}}\rho_{R_{n}^{(2)}}\right)$ in the whole dynamical process, so that the quantity $\sum_{m=1}^{2}\beta_{n}^{(m)}\Delta \widetilde{Q}_{n}^{(m)}$ stays larger than $\Delta \widetilde{S}_{n}$: this means that the Landauer principle still holds although there exist nonzero system-environment correlations.
By contrast, as shown in Fig. \ref{multiviol}(b) and (c), larger values of $\Omega$ (e.g., $\Omega=1.4\omega$, $1.45\omega$) can make $I(\rho_{SR_{n}^{(1)}R_{n}^{(2)}})$ exceed $D\left(\rho_{SR_{n}^{(1)}R_{n}^{(2)}}^{\prime}\parallel\rho_{S_{n}}^{\prime}\rho_{R_{n}^{(1)}}\rho_{R_{n}^{(2)}}\right)$ in some time intervals of the evolution during which the Landauer principle is violated with $\sum_{m=1}^{2}\beta_{n}^{(m)}\Delta \widetilde{Q}_{n}^{(m)}<\Delta \widetilde{S}_{n}$. We notice that, although there are quantitative differences with the case of a single non-Markovian reservoir,   the qualitative results remain the same.

\section{Conclusion}\label{sec4}
In conclusion, we have studied by means of collision models the validity of Landauer principle in a non-Markovian process caused by intracollisions of reservoir ancillary qubits. We have utilized the system-environment correlations formed during the dynamical process to assess the effect of non-Markovianity (memory effects) on Landauer principle. We first consider the situation with the system interacting with a single thermal reservoir. By connecting an exact equality for the system entropy change [see Eq. (\ref{theorem})] to the Landauer principle [see Eq. (\ref{theorem2})], we have reported the condition [see Eq. (\ref{condi})] for its violation in the non-Markovian regime: when the established system-environment correlations, quantified by mutual information, exceed the entropy production then the validity of Landauer principle breaks down. These results have been verified by using the collision model with Heisenberg-like coherent interactions.

We have then generalized the above findings obtained for a single reservoir to the non-equilibrium case when the system is coupled to a composite environment made of $M$ different reservoirs [see Eqs. (\ref{multitheorem} and (\ref{multitheorem2})]. In the presence of these multiple non-Markovian reservoirs, we have proven that when the established correlations between the system and all the reservoirs are larger than the entropy production, the Landauer principle is violated [see. Eq. (\ref{cond-multi})]. We have then verified these results focusing on the simple case of two reservoirs. The findings obtained for the composite environment with multiple reservoirs make it evident how the qualitative behavior concerning the validity of the Landauer principle remains the same as for the case of a single non-Markovian reservoir, with differences emerging only at a quantitative level. Therefore, the complexity of the environment does not appear to have a main role in determining the mechanisms underlying the validity of the Landauer principle.

The results of this work rely on the collision models which are linked to realistic physical models \cite{colli13} and may find experimental realization in all-optical setups\cite{collexp1,collexp2}.

\acknowledgements
In this work Z.X.M. and Y.J.X. are supported by National Natural Science Foundation (China) under Grant Nos.~11574178 and 61675115, and Shandong Provincial Natural Science Foundation (China) under Grant No.~ZR2016JL005. R.L.F. acknowledges Francesco Ciccarello for fruitful discussions.

\appendix

\section{Derivation of Eq. (\ref{DS})} \label{appenA}
In this appendix we report the detailed steps which lead to the expression given in Eq.  (\ref{DS}) of the entropy change $\Delta S_n$ of the system because of the interaction with the reservoir ancilla $R_n$. The derivation is as follows
\begin{eqnarray}
  \Delta S_{n} &=& S(\rho_{S_{n}}^{\prime})-S(\rho_{S_{n}}) \nonumber\\
   &=& S(\rho_{S_{n}}^{\prime})- S(\rho_{SR_{n}})+S(\rho_{R_{n}})-I(\rho_{SR_{n}}) \nonumber\\
   &=& S(\rho_{S_{n}}^{\prime})- S(\rho_{SR_{n}}^{\prime})+S(\rho_{R_{n}})-I(\rho_{SR_{n}}) \nonumber\\
  &=& -\mathrm{Tr}_{S_{n}}\rho_{S_{n}}^{\prime}\ln\rho_{S_{n}}^{\prime}+\mathrm{Tr}\rho_{SR_{n}}^{\prime}\ln\rho_{SR_{n}}^{\prime}
  -\mathrm{Tr}_{R_{n}}\rho_{R_{n}}\ln\rho_{R_{n}} \nonumber\\
  && -I(\rho_{SR_{n}})-\mathrm{Tr}_{R_{n}}\rho_{R_{n}}^{\prime}\ln\rho_{R_{n}}+\mathrm{Tr}_{R_{n}}\rho_{R_{n}}^{\prime}\ln\rho_{R_{n}}\nonumber\\
  &=& -\mathrm{Tr}\rho_{SR_{n}}^{\prime}\ln\rho_{S_{n}}^{\prime}+\mathrm{Tr}\rho_{SR_{n}}^{\prime}\ln\rho_{SR_{n}}^{\prime}-\mathrm{Tr}_{R_{n}}\rho_{R_{n}}
  \ln\rho_{R_{n}} \nonumber\\&&
  -I(\rho_{SR_{n}})-\mathrm{Tr}\rho_{SR_{n}}^{\prime}\ln\rho_{R_{n}}+\mathrm{Tr}_{R_{n}}\rho_{R_{n}}^{\prime}\ln\rho_{R_{n}}\nonumber\\
&=& -\mathrm{Tr}\rho_{SR_{n}}^{\prime}\ln\{\rho_{S_{n}}^{\prime}\rho_{R_{n}}\}+\mathrm{Tr}\rho_{SR_{n}}^{\prime}\ln\rho_{SR_{n}}^{\prime}
-I(\rho_{SR_{n}})\nonumber\\&&+\mathrm{Tr}_{R_{n}}(\rho_{R_{n}}^{\prime}-\rho_{R_{n}})\ln\rho_{R_{n}} \nonumber\\
&=&  D\left(\rho_{SR_{n}}^{\prime}\parallel\rho_{S_{n}}^{\prime}\rho_{R_{n}}\right)
  +\mathrm{Tr}_{R_{n}}(\rho_{R_{n}}^{\prime}-\rho_{R_{n}})\ln\rho_{R_{n}}\nonumber\\
  &&-I(\rho_{SR_{n}}),
\end{eqnarray}

\section{Derivation of Eq. (\ref{multiDS})} \label{appenB}
In this appendix we give the calculations which allow us to obtain the expression of Eq.  (\ref{multiDS}) for the entropy change $\Delta S_n$ of the system due to the interaction with the multiple reservoir ancillas $R^{(1)}_n$, $R^{(2)}_n$, $\ldots$, $R^{(M)}_n$. The derivation is as follows

\begin{widetext}
\begin{eqnarray}
  \Delta S_{n} &=& S\left(\rho_{S_{n}}^{\prime}\right)-S\left(\rho_{S_{n}}\right) \nonumber\\
   &=& S\left(\rho_{S_{n}}^{\prime}\right)-S\left(\rho_{SR_{n}^{(1)}R_{n}^{(2)}...R_{n}^{(M)}}\right)
   +\sum_{m=1}^{M}S\left(\rho_{R_{n}^{(m)}}\right)-I\left(\rho_{SR_{n}^{(1)}R_{n}^{(2)}...R_{n}^{(M)}}\right) \nonumber\\
   &=& S\left(\rho_{S_{n}}^{\prime}\right)- S\left(\rho_{SR_{n}^{(1)}R_{n}^{(2)}...R_{n}^{(M)}}^{\prime}\right)+\sum_{m=1}^{M}S\left(\rho_{R_{n}^{(m)}}\right)
   -I\left(\rho_{SR_{n}^{(1)}R_{n}^{(2)}...R_{n}^{(M)}}\right) \nonumber\\
    &=& -\mathrm{Tr}_{S_{n}}\rho_{S_{n}}^{\prime}\ln\rho_{S_{n}}^{\prime}+\mathrm{Tr}\rho_{SR_{n}^{(1)}R_{n}^{(2)}...R_{n}^{(M)}}^{\prime}
  \ln\rho_{SR_{n}^{(1)}R_{n}^{(2)}...R_{n}^{(M)}}^{\prime}
  -\sum_{m=1}^{M}\mathrm{Tr}_{R_{n}^{(m)}}\rho_{R_{n}^{(m)}}\ln\rho_{R_{n}^{(m)}} \nonumber\\
  && -I\left(\rho_{SR_{n}^{(1)}R_{n}^{(2)}...R_{n}^{(M)}}\right)
  -\sum_{m=1}^{M}\mathrm{Tr}_{R_{n}^{(m)}}\rho_{R_{n}^{(m)}}^{\prime}\ln\rho_{R_{n}^{(m)}}
  +\sum_{m=1}^{M}\mathrm{Tr}_{R_{n}^{(m)}}\rho_{R_{n}^{(m)}}^{\prime}\ln\rho_{R_{n}^{(m)}}\nonumber\\
  &=& -\mathrm{Tr}\rho_{SR_{n}^{(1)}R_{n}^{(2)}...R_{n}^{(M)}}^{\prime}\ln\rho_{S_{n}}^{\prime}
  +\mathrm{Tr}\rho_{SR_{n}^{(1)}R_{n}^{(2)}...R_{n}^{(M)}}^{\prime}\ln\rho_{SR_{n}^{(1)}R_{n}^{(2)}...R_{n}^{(M)}}^{\prime}
  -\sum_{m=1}^{M}\mathrm{Tr}_{R_{n}^{(m)}}\rho_{R_{n}^{(m)}}\ln\rho_{R_{n}^{(m)}} \nonumber\\&&
  -I\left(\rho_{SR_{n}^{(1)}R_{n}^{(2)}...R_{n}^{(M)}}\right)
  -\sum_{m=1}^{M}\mathrm{Tr}\rho_{SR_{n}^{(1)}R_{n}^{(2)}...R_{n}^{(M)}}^{\prime}\ln\rho_{R_{n}^{(m)}}
  +\sum_{m=1}^{M}\mathrm{Tr}_{R_{n}^{(m)}}\rho_{R_{n}^{(m)}}^{\prime}\ln\rho_{R_{n}^{(m)}}\nonumber\\
  &=& -\mathrm{Tr}\rho_{SR_{n}^{(1)}R_{n}^{(2)}...R_{n}^{(M)}}^{\prime}
\ln\left\{\rho_{S_{n}}^{\prime}\prod_{m=1}^{M}\rho_{R_{n}^{(m)}}\right\}
+\mathrm{Tr}\rho_{SR_{n}^{(1)}R_{n}^{(2)}...R_{n}^{(M)}}^{\prime}\ln\rho_{SR_{n}^{(1)}R_{n}^{(2)}...R_{n}^{(M)}}^{\prime}
-I(\rho_{SR_{n}^{(1)}R_{n}^{(2)}...R_{n}^{(M)}})\nonumber\\&&+\sum_{m=1}^{M}\mathrm{Tr}_{R_{n}^{(m)}}(\rho_{R_{n}^{(m)}}^{\prime}-\rho_{R_{n}^{(m)}})
\ln\rho_{R_{n}^{(m)}} \nonumber\\
&=&  D\left(\rho_{SR_{n}^{(1)}R_{n}^{(2)}...R_{n}^{(M)}}^{\prime}\parallel\rho_{S_{n}}^{\prime}\prod_{m=1}^{M}\rho_{R_{n}^{(m)}}\right)
  +\sum_{m=1}^{M}\mathrm{Tr}_{R_{n}^{(m)}}\left(\rho_{R_{n}^{(m)}}^{\prime}-\rho_{R_{n}^{(m)}}\right)
\ln\rho_{R_{n}^{(m)}}-I\left(\rho_{SR_{n}^{(1)}R_{n}^{(2)}...R_{n}^{(M)}}\right).
\end{eqnarray}
\end{widetext}

\end{document}